\documentstyle[HEP93]{article}
\begin{document}
\title{ON THE MEASURE OF SIMPLICIAL QUANTUM GRAVITY\\
       IN FOUR DIMENSIONS \footnotemark[1]}
\author{Wolfgang BEIRL, Harald MARKUM, J\"urgen RIEDLER\\
        \it Institut f\"ur Kernphysik, Technische Universit\"at Wien, A-1040
        Vienna, Austria}
\abstract{\rightskip=1.5pc \leftskip=1.5pc
          We study quantum gravity in the path-integral
          formulation using the Regge calculus. In spite of the unbounded
          gravitational action the existence of an entropy-dominated phase
          is confirmed. The influence of various types of measures on this
          phase structure is investigated and our results are
          compared with those obtained by dynamical triangulation.
          }

\maketitle
\vskip 5mm
\renewcommand{\thefootnote}{\fnsymbol{footnote}}
\footnotetext[1]{Supported in part by "Fonds zur F\"orderung der
wissen\-schaftlichen Forschung" under Contract P9522-PHY.}

{\bf\noindent Introduction and Theory}

The Regge calculus is a useful tool to study non-per\-tur\-bative aspects of
the
quantum-gravity path-integral
\begin{equation} \label{path}
Z = \int D\mu e^{-I_E}
\end{equation}
in a systematic way \cite{hartle,hamber,berg}.
Thereby the continuum Einstein-Hilbert action
\begin{equation}
-I_E=L_P^{-2}\int d^4x\sqrt{g}R-\lambda\int d^4x\sqrt{g}~,
\end{equation}
with $L_P$ the Planck length, $R$ the curvature scalar, $g$ the
determinant of the metric, and $\lambda$ a cosmological constant,
is replaced by the discrete Euclidean action
\begin{equation} \label{act}
-I_E=\beta\sum_t A_t\delta_t-\lambda\sum_s V_s~.
\end{equation}
Triangle areas $A_t$, deficit angles $\delta_t$, and 4-simplex volumes
$V_s$  are calculated from the
squared link lengths $q_l$ given in units of $L_P$ and being the dynamical
quantities \cite{regge}.
The constant $\lambda$ fixes the expectation value of the lattice volume
and the parameter $\beta$ determines the scale.
We define the expectation value of the lattice spacing in units of the
Planck length as
\begin{equation} \label{als}
\ell=(\frac{\beta}{2}\langle q_l\rangle )^{1/2} ~,
\end{equation}
which is an observable rather than a parameter
since the simplicial lattice itself is a quantum object.
Another important observable
is the average curvature measured in units of the average link length.
It is defined as
\begin{equation} \label{ac}
\tilde R = \left(\frac{1}{N_1}\sum_l q_l\right) \frac{\sum_tA_t\delta_t}
{\sum_sV_s}
\end{equation}
with $N_1$ the total number of links.

Obviously, the unpleasant feature of an unbounded gravitational
action is also present
in simplicial quantum gravity, but this does not rule out a priori a
well-defined path integral. As a matter of fact it is possible that
the entropy of the system compensates the unbounded action leading to
the occurence of a 'well-defined' phase as mentioned first by Berg
\cite{berg}.
This can be seen if (\ref{path}) is rewritten as
\begin{equation} \label{Z}
Z=\int_{-\infty}^{+\infty}dI_E~n(I_E)e^{-I_E}~,
\end{equation}
where $n(I_E)$ denotes the state density for a given value $I_E$ of the
action, i.e.~the number of configurations with the same Euclidean
action. If $n(I_E)$ vanishes fast enough for $I_E\rightarrow -\infty$
the integral (\ref{Z}) stays finite in a certain range of $\beta$.
This means that there are
many configurations with small action and only few giving a large
average curvature; the larger the curvature the
smaller its probability.
Configurations with large curvature contain
distorted 4-simplices and are near to leave the Euclidean sector. On
the other hand configurations with almost equilateral simplices
correspond to small average curvatures.

To investigate this mechanism we have set a lower limit
\begin{equation}
\phi_s\ge f\ge 0
\end{equation}
for the fatness of each 4-simplex
\begin{equation}
\phi_s \sim \frac{V_s^2}{\mbox{max}_{l\in s}(q_l^4)}
\end{equation}
restricting the configuration space \cite{chee}.
The gravitational action (\ref{act})
is therefore bounded for a finite number of 4-simplices as long as
$f>0$. The convergence of expectation values in the limit
$f\rightarrow 0$ indeed supports the above entropy hypothesis \cite{wir}.
Since the configuration space grows with smaller values of $f$ an increasing
number of iterations are necessary to reach equilibrium. In order
to make numerical simulations easier a lower limit $f=10^{-5}$ has been
applied to the fatness $\phi_s$ in the following.

We address now the fundamental question about the influence of the
measure on the entropy-dominated phase described above.
A unique definition of the gravitational measure does not exist since it
is not clear which quantities have to be identified with the 'true'
physical degrees of freedom [7-11].
Therefore, we examine different measures of the form
\begin{equation} \label{dmu}
D \mu = ( \prod_l q^{\sigma - 1}_l dq) {\cal F} (q)
\end{equation}
by varying the parameter $\sigma$. This corresponds for $\sigma=0$ to
the scale-invariant measure of Faddeev and Popov \cite{fad} and for
$\sigma=1$ to the uniform measure of deWitt \cite{dewitt}.
The function ${\cal F}$ is equal one for Euclidean configurations and
zero otherwise.

\vskip 5.0mm
{\bf\noindent Results and Discussion}

Computations have been performed on a hypercubic triangulated 4-torus with
$4^4, 6^4$ and $8^4$ vertices. It turned out that finite-volume effects
on one-point functions are small.  The behavior of the average curvature
versus the average lattice spacing is given in Figure~1 for the
$4^4$-vertex system.
The parameter $\sigma$ in (\ref{dmu}) was increased step by step
from $0$ to $1.5$ observing two different regimes.
For $\sigma\le1$ and small lattice spacing,
$0\;^<\!\!\!\!_{-}\;\ell\;^<\!\!\!\!_{\sim}\;0.3$, the
expectation value
$\langle\tilde R\rangle$ is negative and seems to be independent of $\sigma$.
Near the transition point to positive curvature, $\ell\approx 0.4$, the
influence of $\sigma$ becomes more pronounced.
The result for $\sigma=1.5$ differs over the entire range of
$\ell$ from those obtained for $\sigma\le1$.
Even at $\ell=0$ the curvature $\langle\tilde R\rangle$ is
significantly larger for $\sigma > 1$ as illustrated in Figure~2.

\begin{figure}[t]
\setlength{\unitlength}{0.240900pt}
\ifx\plotpoint\undefined\newsavebox{\plotpoint}\fi
\sbox{\plotpoint}{\rule[-0.175pt]{0.350pt}{0.350pt}}%
\begin{picture}(1050,850)(60,000)
\tenrm
\sbox{\plotpoint}{\rule[-0.175pt]{0.350pt}{0.350pt}}%
\put(264,577){\rule[-0.175pt]{173.689pt}{0.350pt}}
\put(264,158){\rule[-0.175pt]{4.818pt}{0.350pt}}
\put(242,158){\makebox(0,0)[r]{-12}}
\put(965,158){\rule[-0.175pt]{4.818pt}{0.350pt}}
\put(264,263){\rule[-0.175pt]{4.818pt}{0.350pt}}
\put(242,263){\makebox(0,0)[r]{-9}}
\put(965,263){\rule[-0.175pt]{4.818pt}{0.350pt}}
\put(264,368){\rule[-0.175pt]{4.818pt}{0.350pt}}
\put(242,368){\makebox(0,0)[r]{-6}}
\put(965,368){\rule[-0.175pt]{4.818pt}{0.350pt}}
\put(264,473){\rule[-0.175pt]{4.818pt}{0.350pt}}
\put(242,473){\makebox(0,0)[r]{-3}}
\put(965,473){\rule[-0.175pt]{4.818pt}{0.350pt}}
\put(264,577){\rule[-0.175pt]{4.818pt}{0.350pt}}
\put(242,577){\makebox(0,0)[r]{0}}
\put(965,577){\rule[-0.175pt]{4.818pt}{0.350pt}}
\put(264,682){\rule[-0.175pt]{4.818pt}{0.350pt}}
\put(242,682){\makebox(0,0)[r]{3}}
\put(965,682){\rule[-0.175pt]{4.818pt}{0.350pt}}
\put(264,787){\rule[-0.175pt]{4.818pt}{0.350pt}}
\put(242,787){\makebox(0,0)[r]{6}}
\put(965,787){\rule[-0.175pt]{4.818pt}{0.350pt}}
\put(264,158){\rule[-0.175pt]{0.350pt}{4.818pt}}
\put(264,113){\makebox(0,0){-0.1}}
\put(264,767){\rule[-0.175pt]{0.350pt}{4.818pt}}
\put(384,158){\rule[-0.175pt]{0.350pt}{4.818pt}}
\put(384,113){\makebox(0,0){0}}
\put(384,767){\rule[-0.175pt]{0.350pt}{4.818pt}}
\put(504,158){\rule[-0.175pt]{0.350pt}{4.818pt}}
\put(504,113){\makebox(0,0){0.1}}
\put(504,767){\rule[-0.175pt]{0.350pt}{4.818pt}}
\put(625,158){\rule[-0.175pt]{0.350pt}{4.818pt}}
\put(625,113){\makebox(0,0){0.2}}
\put(625,767){\rule[-0.175pt]{0.350pt}{4.818pt}}
\put(745,158){\rule[-0.175pt]{0.350pt}{4.818pt}}
\put(745,113){\makebox(0,0){0.3}}
\put(745,767){\rule[-0.175pt]{0.350pt}{4.818pt}}
\put(865,158){\rule[-0.175pt]{0.350pt}{4.818pt}}
\put(865,113){\makebox(0,0){0.4}}
\put(865,767){\rule[-0.175pt]{0.350pt}{4.818pt}}
\put(985,158){\rule[-0.175pt]{0.350pt}{4.818pt}}
\put(985,113){\makebox(0,0){0.5}}
\put(985,767){\rule[-0.175pt]{0.350pt}{4.818pt}}
\put(264,158){\rule[-0.175pt]{173.689pt}{0.350pt}}
\put(985,158){\rule[-0.175pt]{0.350pt}{151.526pt}}
\put(264,787){\rule[-0.175pt]{173.689pt}{0.350pt}}
\put(90,472){\makebox(0,0)[l]{$\langle\tilde R\rangle$}}
\put(624,40){\makebox(0,0){$\ell$}}
\put(264,158){\rule[-0.175pt]{0.350pt}{151.526pt}}
\sbox{\plotpoint}{\rule[-0.250pt]{0.500pt}{0.500pt}}%
\put(380,737){\makebox(0,0)[l]{$\sigma=1.50$}}
\put(384,287){\usebox{\plotpoint}}
\put(384,287){\usebox{\plotpoint}}
\put(403,293){\usebox{\plotpoint}}
\put(423,300){\usebox{\plotpoint}}
\put(443,306){\usebox{\plotpoint}}
\put(462,313){\usebox{\plotpoint}}
\put(482,319){\usebox{\plotpoint}}
\put(502,326){\usebox{\plotpoint}}
\put(521,332){\usebox{\plotpoint}}
\put(541,339){\usebox{\plotpoint}}
\put(561,346){\usebox{\plotpoint}}
\put(580,352){\usebox{\plotpoint}}
\put(599,361){\usebox{\plotpoint}}
\put(614,375){\usebox{\plotpoint}}
\put(630,388){\usebox{\plotpoint}}
\put(646,402){\usebox{\plotpoint}}
\put(661,415){\usebox{\plotpoint}}
\put(677,429){\usebox{\plotpoint}}
\put(692,443){\usebox{\plotpoint}}
\put(707,458){\usebox{\plotpoint}}
\put(722,472){\usebox{\plotpoint}}
\put(737,487){\usebox{\plotpoint}}
\put(752,501){\usebox{\plotpoint}}
\put(761,519){\usebox{\plotpoint}}
\put(768,539){\usebox{\plotpoint}}
\put(775,558){\usebox{\plotpoint}}
\put(782,578){\usebox{\plotpoint}}
\put(789,597){\usebox{\plotpoint}}
\put(796,617){\usebox{\plotpoint}}
\put(803,636){\usebox{\plotpoint}}
\put(810,656){\usebox{\plotpoint}}
\put(817,675){\usebox{\plotpoint}}
\put(824,695){\usebox{\plotpoint}}
\put(826,700){\usebox{\plotpoint}}
\put(328,737){\circle{24}}
\put(384,287){\circle{24}}
\put(594,357){\circle{24}}
\put(684,435){\circle{24}}
\put(757,506){\circle{24}}
\put(826,700){\circle{24}}
\put(380,692){\makebox(0,0)[l]{$\sigma=1.00$}}
\put(384,219){\usebox{\plotpoint}}
\put(384,219){\usebox{\plotpoint}}
\put(404,222){\usebox{\plotpoint}}
\put(424,226){\usebox{\plotpoint}}
\put(445,229){\usebox{\plotpoint}}
\put(465,233){\usebox{\plotpoint}}
\put(486,237){\usebox{\plotpoint}}
\put(506,240){\usebox{\plotpoint}}
\put(527,244){\usebox{\plotpoint}}
\put(547,247){\usebox{\plotpoint}}
\put(567,251){\usebox{\plotpoint}}
\put(588,255){\usebox{\plotpoint}}
\put(608,259){\usebox{\plotpoint}}
\put(628,263){\usebox{\plotpoint}}
\put(649,268){\usebox{\plotpoint}}
\put(669,272){\usebox{\plotpoint}}
\put(689,278){\usebox{\plotpoint}}
\put(709,284){\usebox{\plotpoint}}
\put(728,291){\usebox{\plotpoint}}
\put(745,304){\usebox{\plotpoint}}
\put(761,317){\usebox{\plotpoint}}
\put(777,330){\usebox{\plotpoint}}
\put(793,342){\usebox{\plotpoint}}
\put(810,355){\usebox{\plotpoint}}
\put(826,368){\usebox{\plotpoint}}
\put(843,379){\usebox{\plotpoint}}
\put(862,389){\usebox{\plotpoint}}
\put(878,402){\usebox{\plotpoint}}
\put(888,419){\usebox{\plotpoint}}
\put(899,437){\usebox{\plotpoint}}
\put(910,455){\usebox{\plotpoint}}
\put(919,469){\usebox{\plotpoint}}
\put(328,692){\makebox(0,0){$\spadesuit$}}
\put(384,219){\makebox(0,0){$\spadesuit$}}
\put(582,254){\makebox(0,0){$\spadesuit$}}
\put(663,271){\makebox(0,0){$\spadesuit$}}
\put(727,290){\makebox(0,0){$\spadesuit$}}
\put(832,373){\makebox(0,0){$\spadesuit$}}
\put(875,397){\makebox(0,0){$\spadesuit$}}
\put(919,469){\makebox(0,0){$\spadesuit$}}
\put(380,647){\makebox(0,0)[l]{$\sigma=0.50$}}
\put(384,216){\usebox{\plotpoint}}
\put(384,216){\usebox{\plotpoint}}
\put(404,220){\usebox{\plotpoint}}
\put(424,224){\usebox{\plotpoint}}
\put(445,228){\usebox{\plotpoint}}
\put(465,232){\usebox{\plotpoint}}
\put(485,236){\usebox{\plotpoint}}
\put(506,240){\usebox{\plotpoint}}
\put(526,244){\usebox{\plotpoint}}
\put(546,248){\usebox{\plotpoint}}
\put(567,252){\usebox{\plotpoint}}
\put(587,256){\usebox{\plotpoint}}
\put(607,260){\usebox{\plotpoint}}
\put(628,264){\usebox{\plotpoint}}
\put(648,268){\usebox{\plotpoint}}
\put(669,272){\usebox{\plotpoint}}
\put(689,276){\usebox{\plotpoint}}
\put(709,280){\usebox{\plotpoint}}
\put(730,284){\usebox{\plotpoint}}
\put(750,288){\usebox{\plotpoint}}
\put(770,292){\usebox{\plotpoint}}
\put(791,296){\usebox{\plotpoint}}
\put(811,300){\usebox{\plotpoint}}
\put(830,308){\usebox{\plotpoint}}
\put(849,316){\usebox{\plotpoint}}
\put(867,326){\usebox{\plotpoint}}
\put(870,328){\usebox{\plotpoint}}
\put(328,647){\makebox(0,0){$\diamondsuit$}}
\put(384,216){\makebox(0,0){$\diamondsuit$}}
\put(810,300){\makebox(0,0){$\diamondsuit$}}
\put(851,317){\makebox(0,0){$\diamondsuit$}}
\put(870,328){\makebox(0,0){$\diamondsuit$}}
\put(380,602){\makebox(0,0)[l]{$\sigma=0.25$}}
\put(384,219){\usebox{\plotpoint}}
\put(384,219){\usebox{\plotpoint}}
\put(404,222){\usebox{\plotpoint}}
\put(424,226){\usebox{\plotpoint}}
\put(445,229){\usebox{\plotpoint}}
\put(465,233){\usebox{\plotpoint}}
\put(486,236){\usebox{\plotpoint}}
\put(506,240){\usebox{\plotpoint}}
\put(527,244){\usebox{\plotpoint}}
\put(547,247){\usebox{\plotpoint}}
\put(567,251){\usebox{\plotpoint}}
\put(588,254){\usebox{\plotpoint}}
\put(608,258){\usebox{\plotpoint}}
\put(629,262){\usebox{\plotpoint}}
\put(649,265){\usebox{\plotpoint}}
\put(670,269){\usebox{\plotpoint}}
\put(690,272){\usebox{\plotpoint}}
\put(711,276){\usebox{\plotpoint}}
\put(731,279){\usebox{\plotpoint}}
\put(751,283){\usebox{\plotpoint}}
\put(772,287){\usebox{\plotpoint}}
\put(792,290){\usebox{\plotpoint}}
\put(800,292){\usebox{\plotpoint}}
\put(328,602){\raisebox{-1.2pt}{\makebox(0,0){$\heartsuit$}}}
\put(384,219){\raisebox{-1.2pt}{\makebox(0,0){$\heartsuit$}}}
\put(800,292){\raisebox{-1.2pt}{\makebox(0,0){$\heartsuit$}}}
\put(380,557){\makebox(0,0)[l]{$\sigma=0.0$}}
\put(384,228){\usebox{\plotpoint}}
\put(384,228){\usebox{\plotpoint}}
\put(404,229){\usebox{\plotpoint}}
\put(425,230){\usebox{\plotpoint}}
\put(446,231){\usebox{\plotpoint}}
\put(466,233){\usebox{\plotpoint}}
\put(487,234){\usebox{\plotpoint}}
\put(508,235){\usebox{\plotpoint}}
\put(529,237){\usebox{\plotpoint}}
\put(549,238){\usebox{\plotpoint}}
\put(570,241){\usebox{\plotpoint}}
\put(590,245){\usebox{\plotpoint}}
\put(610,250){\usebox{\plotpoint}}
\put(630,255){\usebox{\plotpoint}}
\put(651,258){\usebox{\plotpoint}}
\put(671,261){\usebox{\plotpoint}}
\put(692,265){\usebox{\plotpoint}}
\put(711,272){\usebox{\plotpoint}}
\put(731,278){\usebox{\plotpoint}}
\put(750,287){\usebox{\plotpoint}}
\put(769,296){\usebox{\plotpoint}}
\put(788,304){\usebox{\plotpoint}}
\put(808,309){\usebox{\plotpoint}}
\put(827,316){\usebox{\plotpoint}}
\put(847,324){\usebox{\plotpoint}}
\put(866,332){\usebox{\plotpoint}}
\put(885,340){\usebox{\plotpoint}}
\put(900,354){\usebox{\plotpoint}}
\put(914,369){\usebox{\plotpoint}}
\put(919,375){\usebox{\plotpoint}}
\put(328,557){\makebox(0,0){$\clubsuit$}}
\put(384,228){\makebox(0,0){$\clubsuit$}}
\put(561,239){\makebox(0,0){$\clubsuit$}}
\put(634,256){\makebox(0,0){$\clubsuit$}}
\put(691,265){\makebox(0,0){$\clubsuit$}}
\put(738,281){\makebox(0,0){$\clubsuit$}}
\put(780,302){\makebox(0,0){$\clubsuit$}}
\put(819,313){\makebox(0,0){$\clubsuit$}}
\put(854,327){\makebox(0,0){$\clubsuit$}}
\put(887,341){\makebox(0,0){$\clubsuit$}}
\put(919,375){\makebox(0,0){$\clubsuit$}}
\end{picture}
{ \small
  \hspace{0.2cm}
  Figure~1: Average curvature $\langle \tilde{R} \rangle$
  as a function of the lattice spacing $\ell$ within the Regge approach for
  different types of the measure parametrized by $\sigma \geq 0$.
  The behavior of $\langle \tilde{R} \rangle$
  in the region of small $\ell$ is almost independent of $\sigma$
  as long as $\sigma \leq 1$.
  }
\end{figure}

\begin{figure}
\setlength{\unitlength}{0.240900pt}
\ifx\plotpoint\undefined\newsavebox{\plotpoint}\fi
\sbox{\plotpoint}{\rule[-0.175pt]{0.350pt}{0.350pt}}%
\begin{picture}(1050,850)(56,000)
\tenrm
\sbox{\plotpoint}{\rule[-0.175pt]{0.350pt}{0.350pt}}%
\put(264,158){\rule[-0.175pt]{4.818pt}{0.350pt}}
\put(242,158){\makebox(0,0)[r]{-11}}
\put(965,158){\rule[-0.175pt]{4.818pt}{0.350pt}}
\put(264,237){\rule[-0.175pt]{4.818pt}{0.350pt}}
\put(242,237){\makebox(0,0)[r]{-10.5}}
\put(965,237){\rule[-0.175pt]{4.818pt}{0.350pt}}
\put(264,315){\rule[-0.175pt]{4.818pt}{0.350pt}}
\put(242,315){\makebox(0,0)[r]{-10}}
\put(965,315){\rule[-0.175pt]{4.818pt}{0.350pt}}
\put(264,394){\rule[-0.175pt]{4.818pt}{0.350pt}}
\put(242,394){\makebox(0,0)[r]{-9.5}}
\put(965,394){\rule[-0.175pt]{4.818pt}{0.350pt}}
\put(264,473){\rule[-0.175pt]{4.818pt}{0.350pt}}
\put(242,473){\makebox(0,0)[r]{-9}}
\put(965,473){\rule[-0.175pt]{4.818pt}{0.350pt}}
\put(264,551){\rule[-0.175pt]{4.818pt}{0.350pt}}
\put(242,551){\makebox(0,0)[r]{-8.5}}
\put(965,551){\rule[-0.175pt]{4.818pt}{0.350pt}}
\put(264,630){\rule[-0.175pt]{4.818pt}{0.350pt}}
\put(242,630){\makebox(0,0)[r]{-8}}
\put(965,630){\rule[-0.175pt]{4.818pt}{0.350pt}}
\put(264,708){\rule[-0.175pt]{4.818pt}{0.350pt}}
\put(242,708){\makebox(0,0)[r]{-7.5}}
\put(965,708){\rule[-0.175pt]{4.818pt}{0.350pt}}
\put(264,787){\rule[-0.175pt]{4.818pt}{0.350pt}}
\put(242,787){\makebox(0,0)[r]{-7}}
\put(965,787){\rule[-0.175pt]{4.818pt}{0.350pt}}
\put(306,158){\rule[-0.175pt]{0.350pt}{4.818pt}}
\put(306,113){\makebox(0,0){0}}
\put(306,767){\rule[-0.175pt]{0.350pt}{4.818pt}}
\put(391,158){\rule[-0.175pt]{0.350pt}{4.818pt}}
\put(391,113){\makebox(0,0){0.2}}
\put(391,767){\rule[-0.175pt]{0.350pt}{4.818pt}}
\put(476,158){\rule[-0.175pt]{0.350pt}{4.818pt}}
\put(476,113){\makebox(0,0){0.4}}
\put(476,767){\rule[-0.175pt]{0.350pt}{4.818pt}}
\put(561,158){\rule[-0.175pt]{0.350pt}{4.818pt}}
\put(561,113){\makebox(0,0){0.6}}
\put(561,767){\rule[-0.175pt]{0.350pt}{4.818pt}}
\put(646,158){\rule[-0.175pt]{0.350pt}{4.818pt}}
\put(646,113){\makebox(0,0){0.8}}
\put(646,767){\rule[-0.175pt]{0.350pt}{4.818pt}}
\put(731,158){\rule[-0.175pt]{0.350pt}{4.818pt}}
\put(731,113){\makebox(0,0){1}}
\put(731,767){\rule[-0.175pt]{0.350pt}{4.818pt}}
\put(815,158){\rule[-0.175pt]{0.350pt}{4.818pt}}
\put(815,113){\makebox(0,0){1.2}}
\put(815,767){\rule[-0.175pt]{0.350pt}{4.818pt}}
\put(900,158){\rule[-0.175pt]{0.350pt}{4.818pt}}
\put(900,113){\makebox(0,0){1.4}}
\put(900,767){\rule[-0.175pt]{0.350pt}{4.818pt}}
\put(985,158){\rule[-0.175pt]{0.350pt}{4.818pt}}
\put(985,113){\makebox(0,0){1.6}}
\put(985,767){\rule[-0.175pt]{0.350pt}{4.818pt}}
\put(264,158){\rule[-0.175pt]{173.689pt}{0.350pt}}
\put(985,158){\rule[-0.175pt]{0.350pt}{151.526pt}}
\put(264,787){\rule[-0.175pt]{173.689pt}{0.350pt}}
\put(80,472){\makebox(0,0)[l]{$\langle\tilde R\rangle$}}
\put(624,40){\makebox(0,0){$\sigma$}}
\put(264,158){\rule[-0.175pt]{0.350pt}{151.526pt}}
\sbox{\plotpoint}{\rule[-0.250pt]{0.500pt}{0.500pt}}%
\put(380,708){\makebox(0,0)[l]{$\ell=0$}}
\put(306,317){\usebox{\plotpoint}}
\put(306,317){\usebox{\plotpoint}}
\put(325,309){\usebox{\plotpoint}}
\put(344,301){\usebox{\plotpoint}}
\put(363,293){\usebox{\plotpoint}}
\put(382,285){\usebox{\plotpoint}}
\put(402,277){\usebox{\plotpoint}}
\put(422,272){\usebox{\plotpoint}}
\put(442,270){\usebox{\plotpoint}}
\put(463,268){\usebox{\plotpoint}}
\put(484,266){\usebox{\plotpoint}}
\put(504,264){\usebox{\plotpoint}}
\put(525,263){\usebox{\plotpoint}}
\put(546,264){\usebox{\plotpoint}}
\put(566,265){\usebox{\plotpoint}}
\put(587,267){\usebox{\plotpoint}}
\put(608,268){\usebox{\plotpoint}}
\put(628,269){\usebox{\plotpoint}}
\put(649,271){\usebox{\plotpoint}}
\put(670,272){\usebox{\plotpoint}}
\put(691,273){\usebox{\plotpoint}}
\put(711,274){\usebox{\plotpoint}}
\put(731,277){\usebox{\plotpoint}}
\put(745,293){\usebox{\plotpoint}}
\put(758,309){\usebox{\plotpoint}}
\put(771,325){\usebox{\plotpoint}}
\put(784,341){\usebox{\plotpoint}}
\put(797,357){\usebox{\plotpoint}}
\put(811,373){\usebox{\plotpoint}}
\put(824,389){\usebox{\plotpoint}}
\put(837,405){\usebox{\plotpoint}}
\put(848,423){\usebox{\plotpoint}}
\put(858,441){\usebox{\plotpoint}}
\put(869,458){\usebox{\plotpoint}}
\put(880,476){\usebox{\plotpoint}}
\put(891,494){\usebox{\plotpoint}}
\put(901,512){\usebox{\plotpoint}}
\put(912,529){\usebox{\plotpoint}}
\put(923,547){\usebox{\plotpoint}}
\put(934,565){\usebox{\plotpoint}}
\put(943,580){\usebox{\plotpoint}}
\put(306,317){\makebox(0,0){$\clubsuit$}}
\put(412,274){\raisebox{-1.2pt}{\makebox(0,0){$\heartsuit$}}}
\put(518,263){\makebox(0,0){$\diamondsuit$}}
\put(731,276){\makebox(0,0){$\spadesuit$}}
\put(837,405){\makebox(0,0){$\triangle$}}
\put(943,580){\circle{24}}
\end{picture}
{ \small
  \hspace{0.2cm}
  Figure~2: Average curvature $\langle \tilde{R} \rangle$ versus
  measure parameter $\sigma$ in the case of pure entropy,
  $\ell = 0 \leftrightarrow \beta = 0$.
  For $0\le\sigma\le 1$ $\langle \tilde{R} \rangle$ stays rather
  constant whereas for $\sigma > 1$ significant deviations are observed.
  }
\end{figure}

To yield more information about the lattice geometry the behavior of
the triangle areas and deficit angles is examined separately.
Figure~3 displays the scale-invariant quantity
$\langle A_t \rangle / \langle q_l \rangle$
as a function of $\ell$ for different
values of $\sigma$. For small lattice spacing,
$0\;^<\!\!\!\!_{-}\;\ell\;^<\!\!\!\!_{\sim}\;0.4$,
this ratio stays almost constant and somewhat below the value
$\frac{\sqrt{3}}{4} \approx 0.433$ corresponding to equilateral triangles.
The curve for $\sigma = 1.5$ lies significantly
below the others. Across the transition at $\ell \approx 0.45$ the
ratio $\langle A_t \rangle / \langle q_l \rangle$
decreases indicating a distortion of the triangles.

\setlength{\unitlength}{0.240900pt}
\ifx\plotpoint\undefined\newsavebox{\plotpoint}\fi
\sbox{\plotpoint}{\rule[-0.175pt]{0.350pt}{0.350pt}}%
\begin{picture}(1050,850)(115,0)
\tenrm
\sbox{\plotpoint}{\rule[-0.175pt]{0.350pt}{0.350pt}}%
\put(264,158){\rule[-0.175pt]{4.818pt}{0.350pt}}
\put(242,158){\makebox(0,0)[r]{0.3}}
\put(965,158){\rule[-0.175pt]{4.818pt}{0.350pt}}
\put(264,263){\rule[-0.175pt]{4.818pt}{0.350pt}}
\put(242,263){\makebox(0,0)[r]{0.325}}
\put(965,263){\rule[-0.175pt]{4.818pt}{0.350pt}}
\put(264,368){\rule[-0.175pt]{4.818pt}{0.350pt}}
\put(242,368){\makebox(0,0)[r]{0.35}}
\put(965,368){\rule[-0.175pt]{4.818pt}{0.350pt}}
\put(264,473){\rule[-0.175pt]{4.818pt}{0.350pt}}
\put(242,473){\makebox(0,0)[r]{0.375}}
\put(965,473){\rule[-0.175pt]{4.818pt}{0.350pt}}
\put(264,577){\rule[-0.175pt]{4.818pt}{0.350pt}}
\put(242,577){\makebox(0,0)[r]{0.4}}
\put(965,577){\rule[-0.175pt]{4.818pt}{0.350pt}}
\put(264,682){\rule[-0.175pt]{4.818pt}{0.350pt}}
\put(242,682){\makebox(0,0)[r]{0.425}}
\put(965,682){\rule[-0.175pt]{4.818pt}{0.350pt}}
\put(264,787){\rule[-0.175pt]{4.818pt}{0.350pt}}
\put(242,787){\makebox(0,0)[r]{0.45}}
\put(965,787){\rule[-0.175pt]{4.818pt}{0.350pt}}
\put(264,158){\rule[-0.175pt]{0.350pt}{4.818pt}}
\put(264,113){\makebox(0,0){-0.1}}
\put(264,767){\rule[-0.175pt]{0.350pt}{4.818pt}}
\put(384,158){\rule[-0.175pt]{0.350pt}{4.818pt}}
\put(384,113){\makebox(0,0){0}}
\put(384,767){\rule[-0.175pt]{0.350pt}{4.818pt}}
\put(504,158){\rule[-0.175pt]{0.350pt}{4.818pt}}
\put(504,113){\makebox(0,0){0.1}}
\put(504,767){\rule[-0.175pt]{0.350pt}{4.818pt}}
\put(625,158){\rule[-0.175pt]{0.350pt}{4.818pt}}
\put(625,113){\makebox(0,0){0.2}}
\put(625,767){\rule[-0.175pt]{0.350pt}{4.818pt}}
\put(745,158){\rule[-0.175pt]{0.350pt}{4.818pt}}
\put(745,113){\makebox(0,0){0.3}}
\put(745,767){\rule[-0.175pt]{0.350pt}{4.818pt}}
\put(865,158){\rule[-0.175pt]{0.350pt}{4.818pt}}
\put(865,113){\makebox(0,0){0.4}}
\put(865,767){\rule[-0.175pt]{0.350pt}{4.818pt}}
\put(985,158){\rule[-0.175pt]{0.350pt}{4.818pt}}
\put(985,113){\makebox(0,0){0.5}}
\put(985,767){\rule[-0.175pt]{0.350pt}{4.818pt}}
\put(264,158){\rule[-0.175pt]{173.689pt}{0.350pt}}
\put(985,158){\rule[-0.175pt]{0.350pt}{151.526pt}}
\put(264,787){\rule[-0.175pt]{173.689pt}{0.350pt}}
\put(75,577){\makebox(0,0)[l]
            {\shortstack{$\frac{\langle A_t\rangle}{\langle q_\ell\rangle}$}}}
\put(624,40){\makebox(0,0){$\ell$}}
\put(264,158){\rule[-0.175pt]{0.350pt}{151.526pt}}
\sbox{\plotpoint}{\rule[-0.250pt]{0.500pt}{0.500pt}}%
\put(380,745){\makebox(0,0)[l]{$\sigma=1.50$}}
\put(384,405){\usebox{\plotpoint}}
\put(384,405){\usebox{\plotpoint}}
\put(404,402){\usebox{\plotpoint}}
\put(425,400){\usebox{\plotpoint}}
\put(445,398){\usebox{\plotpoint}}
\put(466,396){\usebox{\plotpoint}}
\put(487,394){\usebox{\plotpoint}}
\put(507,392){\usebox{\plotpoint}}
\put(528,389){\usebox{\plotpoint}}
\put(549,387){\usebox{\plotpoint}}
\put(569,385){\usebox{\plotpoint}}
\put(590,383){\usebox{\plotpoint}}
\put(610,379){\usebox{\plotpoint}}
\put(631,376){\usebox{\plotpoint}}
\put(651,372){\usebox{\plotpoint}}
\put(672,369){\usebox{\plotpoint}}
\put(692,364){\usebox{\plotpoint}}
\put(712,359){\usebox{\plotpoint}}
\put(732,354){\usebox{\plotpoint}}
\put(752,349){\usebox{\plotpoint}}
\put(771,339){\usebox{\plotpoint}}
\put(789,329){\usebox{\plotpoint}}
\put(807,318){\usebox{\plotpoint}}
\put(825,308){\usebox{\plotpoint}}
\put(826,308){\usebox{\plotpoint}}
\put(328,745){\circle{24}}
\put(384,405){\circle{24}}
\put(594,383){\circle{24}}
\put(684,367){\circle{24}}
\put(757,348){\circle{24}}
\put(826,308){\circle{24}}
\put(380,700){\makebox(0,0)[l]{$\sigma=1.00$}}
\put(384,471){\usebox{\plotpoint}}
\put(384,471){\usebox{\plotpoint}}
\put(404,470){\usebox{\plotpoint}}
\put(425,469){\usebox{\plotpoint}}
\put(446,468){\usebox{\plotpoint}}
\put(466,467){\usebox{\plotpoint}}
\put(487,466){\usebox{\plotpoint}}
\put(508,465){\usebox{\plotpoint}}
\put(529,465){\usebox{\plotpoint}}
\put(549,464){\usebox{\plotpoint}}
\put(570,463){\usebox{\plotpoint}}
\put(591,463){\usebox{\plotpoint}}
\put(612,463){\usebox{\plotpoint}}
\put(632,463){\usebox{\plotpoint}}
\put(653,463){\usebox{\plotpoint}}
\put(674,462){\usebox{\plotpoint}}
\put(695,460){\usebox{\plotpoint}}
\put(715,458){\usebox{\plotpoint}}
\put(736,455){\usebox{\plotpoint}}
\put(756,451){\usebox{\plotpoint}}
\put(776,447){\usebox{\plotpoint}}
\put(797,442){\usebox{\plotpoint}}
\put(817,438){\usebox{\plotpoint}}
\put(837,435){\usebox{\plotpoint}}
\put(858,435){\usebox{\plotpoint}}
\put(878,433){\usebox{\plotpoint}}
\put(897,425){\usebox{\plotpoint}}
\put(917,416){\usebox{\plotpoint}}
\put(919,416){\usebox{\plotpoint}}
\put(328,700){\makebox(0,0){$\spadesuit$}}
\put(384,471){\makebox(0,0){$\spadesuit$}}
\put(582,463){\makebox(0,0){$\spadesuit$}}
\put(663,463){\makebox(0,0){$\spadesuit$}}
\put(727,458){\makebox(0,0){$\spadesuit$}}
\put(832,435){\makebox(0,0){$\spadesuit$}}
\put(875,435){\makebox(0,0){$\spadesuit$}}
\put(919,416){\makebox(0,0){$\spadesuit$}}
\put(380,655){\makebox(0,0)[l]{$\sigma=0.50$}}
\put(384,505){\usebox{\plotpoint}}
\put(384,505){\usebox{\plotpoint}}
\put(404,504){\usebox{\plotpoint}}
\put(425,503){\usebox{\plotpoint}}
\put(446,502){\usebox{\plotpoint}}
\put(466,502){\usebox{\plotpoint}}
\put(487,501){\usebox{\plotpoint}}
\put(508,500){\usebox{\plotpoint}}
\put(529,499){\usebox{\plotpoint}}
\put(549,499){\usebox{\plotpoint}}
\put(570,498){\usebox{\plotpoint}}
\put(591,497){\usebox{\plotpoint}}
\put(612,496){\usebox{\plotpoint}}
\put(632,496){\usebox{\plotpoint}}
\put(653,495){\usebox{\plotpoint}}
\put(674,494){\usebox{\plotpoint}}
\put(695,494){\usebox{\plotpoint}}
\put(715,493){\usebox{\plotpoint}}
\put(736,492){\usebox{\plotpoint}}
\put(757,491){\usebox{\plotpoint}}
\put(778,491){\usebox{\plotpoint}}
\put(798,490){\usebox{\plotpoint}}
\put(819,489){\usebox{\plotpoint}}
\put(840,487){\usebox{\plotpoint}}
\put(860,485){\usebox{\plotpoint}}
\put(870,484){\usebox{\plotpoint}}
\put(328,655){\makebox(0,0){$\diamondsuit$}}
\put(384,505){\makebox(0,0){$\diamondsuit$}}
\put(810,490){\makebox(0,0){$\diamondsuit$}}
\put(851,487){\makebox(0,0){$\diamondsuit$}}
\put(870,484){\makebox(0,0){$\diamondsuit$}}
\put(380,260){\makebox(0,0)[l]{$\sigma=0.25$}}
\put(384,518){\usebox{\plotpoint}}
\put(384,518){\usebox{\plotpoint}}
\put(404,517){\usebox{\plotpoint}}
\put(425,516){\usebox{\plotpoint}}
\put(446,516){\usebox{\plotpoint}}
\put(466,515){\usebox{\plotpoint}}
\put(487,514){\usebox{\plotpoint}}
\put(508,514){\usebox{\plotpoint}}
\put(529,513){\usebox{\plotpoint}}
\put(549,512){\usebox{\plotpoint}}
\put(570,512){\usebox{\plotpoint}}
\put(591,511){\usebox{\plotpoint}}
\put(612,510){\usebox{\plotpoint}}
\put(632,510){\usebox{\plotpoint}}
\put(653,509){\usebox{\plotpoint}}
\put(674,508){\usebox{\plotpoint}}
\put(695,508){\usebox{\plotpoint}}
\put(715,507){\usebox{\plotpoint}}
\put(736,506){\usebox{\plotpoint}}
\put(757,506){\usebox{\plotpoint}}
\put(778,505){\usebox{\plotpoint}}
\put(798,505){\usebox{\plotpoint}}
\put(819,503){\usebox{\plotpoint}}
\put(840,500){\usebox{\plotpoint}}
\put(860,498){\usebox{\plotpoint}}
\put(881,496){\usebox{\plotpoint}}
\put(890,496){\usebox{\plotpoint}}
\put(328,260){\raisebox{-1.2pt}{\makebox(0,0){$\heartsuit$}}}
\put(384,518){\raisebox{-1.2pt}{\makebox(0,0){$\heartsuit$}}}
\put(800,505){\raisebox{-1.2pt}{\makebox(0,0){$\heartsuit$}}}
\put(890,496){\raisebox{-1.2pt}{\makebox(0,0){$\heartsuit$}}}
\put(380,215){\makebox(0,0)[l]{$\sigma=0.0$}}
\put(384,527){\usebox{\plotpoint}}
\put(384,527){\usebox{\plotpoint}}
\put(404,526){\usebox{\plotpoint}}
\put(425,526){\usebox{\plotpoint}}
\put(446,526){\usebox{\plotpoint}}
\put(467,526){\usebox{\plotpoint}}
\put(487,526){\usebox{\plotpoint}}
\put(508,526){\usebox{\plotpoint}}
\put(529,526){\usebox{\plotpoint}}
\put(550,526){\usebox{\plotpoint}}
\put(570,525){\usebox{\plotpoint}}
\put(591,524){\usebox{\plotpoint}}
\put(612,523){\usebox{\plotpoint}}
\put(633,523){\usebox{\plotpoint}}
\put(653,522){\usebox{\plotpoint}}
\put(674,522){\usebox{\plotpoint}}
\put(695,521){\usebox{\plotpoint}}
\put(715,520){\usebox{\plotpoint}}
\put(736,519){\usebox{\plotpoint}}
\put(757,516){\usebox{\plotpoint}}
\put(777,513){\usebox{\plotpoint}}
\put(798,512){\usebox{\plotpoint}}
\put(819,510){\usebox{\plotpoint}}
\put(839,508){\usebox{\plotpoint}}
\put(860,506){\usebox{\plotpoint}}
\put(881,504){\usebox{\plotpoint}}
\put(900,497){\usebox{\plotpoint}}
\put(919,489){\usebox{\plotpoint}}
\put(328,215){\makebox(0,0){$\clubsuit$}}
\put(384,527){\makebox(0,0){$\clubsuit$}}
\put(561,526){\makebox(0,0){$\clubsuit$}}
\put(634,523){\makebox(0,0){$\clubsuit$}}
\put(691,522){\makebox(0,0){$\clubsuit$}}
\put(738,519){\makebox(0,0){$\clubsuit$}}
\put(780,513){\makebox(0,0){$\clubsuit$}}
\put(819,511){\makebox(0,0){$\clubsuit$}}
\put(854,507){\makebox(0,0){$\clubsuit$}}
\put(887,504){\makebox(0,0){$\clubsuit$}}
\put(919,489){\makebox(0,0){$\clubsuit$}}
\end{picture}
{ \small
  \hspace{0.2cm}
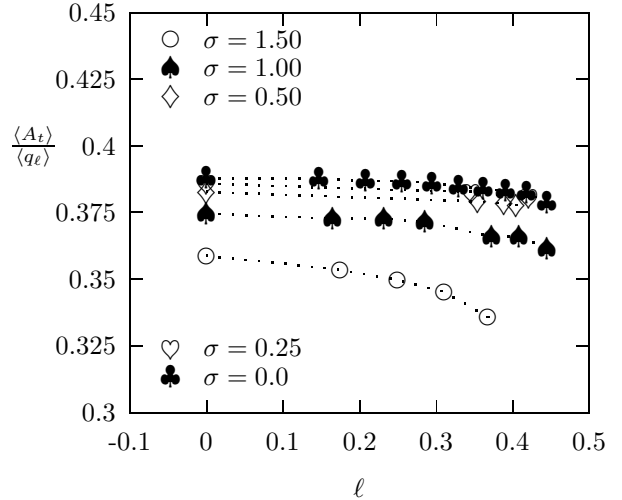
  Figure~3: Ratio $\langle A_t \rangle / \langle q_l \rangle$ as a
  function of the lattice spacing $\ell$ for different measures parametrized
  by $\sigma \geq 0$. In the well-defined phase this
  scale-invariant quantity stays almost constant and
  somewhat below the maximum value $\frac{\sqrt{3}}{4} \approx 0.433$ for
  equilateral triangles. The expectation values decrease slightly with
  increasing $\sigma$.
}

The expectation value of the average deficit angle
$\langle \delta_t \rangle$ as a function of $\ell$ is depicted in Figure~4.
Surprisingly, $\langle \delta_t \rangle$ stays negative
even above the transition to large
positive curvature. In the well-defined phase the absolute
value of $\langle \delta_t \rangle$ is rather small
compared to $\pi$. This can be understood by considering
the hypercubic triangulation of the
4-torus that contains two different types of triangles. The number
of 4-simplices sharing a triangle of the first type is $6$
while it is $4$ for the second type.
The contributions of these two types
to the average deficit angle almost cancel each other leaving
a small negative value \cite{berg}.

\begin{figure}[t]
\setlength{\unitlength}{0.240900pt}
\ifx\plotpoint\undefined\newsavebox{\plotpoint}\fi
\sbox{\plotpoint}{\rule[-0.175pt]{0.350pt}{0.350pt}}%
\begin{picture}(1050,850)(60,0)
\tenrm
\sbox{\plotpoint}{\rule[-0.175pt]{0.350pt}{0.350pt}}%
\put(264,661){\rule[-0.175pt]{173.689pt}{0.350pt}}
\put(264,158){\rule[-0.175pt]{4.818pt}{0.350pt}}
\put(242,158){\makebox(0,0)[r]{-0.1}}
\put(965,158){\rule[-0.175pt]{4.818pt}{0.350pt}}
\put(264,284){\rule[-0.175pt]{4.818pt}{0.350pt}}
\put(242,284){\makebox(0,0)[r]{-0.075}}
\put(965,284){\rule[-0.175pt]{4.818pt}{0.350pt}}
\put(264,410){\rule[-0.175pt]{4.818pt}{0.350pt}}
\put(242,410){\makebox(0,0)[r]{-0.05}}
\put(965,410){\rule[-0.175pt]{4.818pt}{0.350pt}}
\put(264,535){\rule[-0.175pt]{4.818pt}{0.350pt}}
\put(242,535){\makebox(0,0)[r]{-0.025}}
\put(965,535){\rule[-0.175pt]{4.818pt}{0.350pt}}
\put(264,661){\rule[-0.175pt]{4.818pt}{0.350pt}}
\put(242,661){\makebox(0,0)[r]{0}}
\put(965,661){\rule[-0.175pt]{4.818pt}{0.350pt}}
\put(264,787){\rule[-0.175pt]{4.818pt}{0.350pt}}
\put(242,787){\makebox(0,0)[r]{0.025}}
\put(965,787){\rule[-0.175pt]{4.818pt}{0.350pt}}
\put(264,158){\rule[-0.175pt]{0.350pt}{4.818pt}}
\put(264,113){\makebox(0,0){-0.1}}
\put(264,767){\rule[-0.175pt]{0.350pt}{4.818pt}}
\put(384,158){\rule[-0.175pt]{0.350pt}{4.818pt}}
\put(384,113){\makebox(0,0){0}}
\put(384,767){\rule[-0.175pt]{0.350pt}{4.818pt}}
\put(504,158){\rule[-0.175pt]{0.350pt}{4.818pt}}
\put(504,113){\makebox(0,0){0.1}}
\put(504,767){\rule[-0.175pt]{0.350pt}{4.818pt}}
\put(625,158){\rule[-0.175pt]{0.350pt}{4.818pt}}
\put(625,113){\makebox(0,0){0.2}}
\put(625,767){\rule[-0.175pt]{0.350pt}{4.818pt}}
\put(745,158){\rule[-0.175pt]{0.350pt}{4.818pt}}
\put(745,113){\makebox(0,0){0.3}}
\put(745,767){\rule[-0.175pt]{0.350pt}{4.818pt}}
\put(865,158){\rule[-0.175pt]{0.350pt}{4.818pt}}
\put(865,113){\makebox(0,0){0.4}}
\put(865,767){\rule[-0.175pt]{0.350pt}{4.818pt}}
\put(985,158){\rule[-0.175pt]{0.350pt}{4.818pt}}
\put(985,113){\makebox(0,0){0.5}}
\put(985,767){\rule[-0.175pt]{0.350pt}{4.818pt}}
\put(264,158){\rule[-0.175pt]{173.689pt}{0.350pt}}
\put(985,158){\rule[-0.175pt]{0.350pt}{151.526pt}}
\put(264,787){\rule[-0.175pt]{173.689pt}{0.350pt}}
\put(85,661){\makebox(0,0)[l]{\shortstack{$\langle\delta_t\rangle$}}}
\put(624,40){\makebox(0,0){$\ell$}}
\put(264,158){\rule[-0.175pt]{0.350pt}{151.526pt}}
\sbox{\plotpoint}{\rule[-0.250pt]{0.500pt}{0.500pt}}%
\put(380,400){\makebox(0,0)[l]{$\sigma=1.50$}}
\put(384,585){\usebox{\plotpoint}}
\put(384,585){\usebox{\plotpoint}}
\put(404,579){\usebox{\plotpoint}}
\put(424,574){\usebox{\plotpoint}}
\put(444,569){\usebox{\plotpoint}}
\put(464,563){\usebox{\plotpoint}}
\put(484,558){\usebox{\plotpoint}}
\put(504,553){\usebox{\plotpoint}}
\put(524,548){\usebox{\plotpoint}}
\put(544,542){\usebox{\plotpoint}}
\put(564,537){\usebox{\plotpoint}}
\put(584,532){\usebox{\plotpoint}}
\put(603,523){\usebox{\plotpoint}}
\put(620,511){\usebox{\plotpoint}}
\put(637,499){\usebox{\plotpoint}}
\put(654,487){\usebox{\plotpoint}}
\put(671,475){\usebox{\plotpoint}}
\put(688,463){\usebox{\plotpoint}}
\put(704,451){\usebox{\plotpoint}}
\put(721,439){\usebox{\plotpoint}}
\put(738,426){\usebox{\plotpoint}}
\put(754,414){\usebox{\plotpoint}}
\put(764,396){\usebox{\plotpoint}}
\put(773,378){\usebox{\plotpoint}}
\put(783,359){\usebox{\plotpoint}}
\put(792,340){\usebox{\plotpoint}}
\put(801,322){\usebox{\plotpoint}}
\put(810,303){\usebox{\plotpoint}}
\put(819,284){\usebox{\plotpoint}}
\put(826,271){\usebox{\plotpoint}}
\put(328,400){\circle{24}}
\put(384,585){\circle{24}}
\put(594,530){\circle{24}}
\put(684,467){\circle{24}}
\put(757,413){\circle{24}}
\put(826,271){\circle{24}}
\put(380,355){\makebox(0,0)[l]{$\sigma=1.00$}}
\put(384,625){\usebox{\plotpoint}}
\put(384,625){\usebox{\plotpoint}}
\put(404,622){\usebox{\plotpoint}}
\put(425,619){\usebox{\plotpoint}}
\put(445,616){\usebox{\plotpoint}}
\put(466,614){\usebox{\plotpoint}}
\put(486,611){\usebox{\plotpoint}}
\put(507,608){\usebox{\plotpoint}}
\put(528,606){\usebox{\plotpoint}}
\put(548,603){\usebox{\plotpoint}}
\put(569,600){\usebox{\plotpoint}}
\put(589,598){\usebox{\plotpoint}}
\put(610,596){\usebox{\plotpoint}}
\put(631,594){\usebox{\plotpoint}}
\put(651,592){\usebox{\plotpoint}}
\put(672,590){\usebox{\plotpoint}}
\put(692,585){\usebox{\plotpoint}}
\put(713,581){\usebox{\plotpoint}}
\put(732,575){\usebox{\plotpoint}}
\put(750,564){\usebox{\plotpoint}}
\put(767,553){\usebox{\plotpoint}}
\put(784,541){\usebox{\plotpoint}}
\put(802,530){\usebox{\plotpoint}}
\put(819,519){\usebox{\plotpoint}}
\put(837,509){\usebox{\plotpoint}}
\put(856,500){\usebox{\plotpoint}}
\put(875,492){\usebox{\plotpoint}}
\put(888,475){\usebox{\plotpoint}}
\put(900,459){\usebox{\plotpoint}}
\put(913,442){\usebox{\plotpoint}}
\put(919,436){\usebox{\plotpoint}}
\put(328,355){\makebox(0,0){$\spadesuit$}}
\put(384,625){\makebox(0,0){$\spadesuit$}}
\put(582,599){\makebox(0,0){$\spadesuit$}}
\put(663,592){\makebox(0,0){$\spadesuit$}}
\put(727,579){\makebox(0,0){$\spadesuit$}}
\put(832,512){\makebox(0,0){$\spadesuit$}}
\put(875,493){\makebox(0,0){$\spadesuit$}}
\put(919,436){\makebox(0,0){$\spadesuit$}}
\put(380,310){\makebox(0,0)[l]{$\sigma=0.50$}}
\put(384,609){\usebox{\plotpoint}}
\put(384,609){\usebox{\plotpoint}}
\put(404,606){\usebox{\plotpoint}}
\put(425,603){\usebox{\plotpoint}}
\put(445,601){\usebox{\plotpoint}}
\put(466,598){\usebox{\plotpoint}}
\put(486,595){\usebox{\plotpoint}}
\put(507,593){\usebox{\plotpoint}}
\put(528,590){\usebox{\plotpoint}}
\put(548,587){\usebox{\plotpoint}}
\put(569,585){\usebox{\plotpoint}}
\put(589,582){\usebox{\plotpoint}}
\put(610,579){\usebox{\plotpoint}}
\put(631,577){\usebox{\plotpoint}}
\put(651,574){\usebox{\plotpoint}}
\put(672,571){\usebox{\plotpoint}}
\put(692,569){\usebox{\plotpoint}}
\put(713,566){\usebox{\plotpoint}}
\put(733,563){\usebox{\plotpoint}}
\put(754,561){\usebox{\plotpoint}}
\put(775,558){\usebox{\plotpoint}}
\put(795,555){\usebox{\plotpoint}}
\put(816,552){\usebox{\plotpoint}}
\put(835,546){\usebox{\plotpoint}}
\put(855,540){\usebox{\plotpoint}}
\put(870,534){\usebox{\plotpoint}}
\put(328,310){\makebox(0,0){$\diamondsuit$}}
\put(384,609){\makebox(0,0){$\diamondsuit$}}
\put(810,554){\makebox(0,0){$\diamondsuit$}}
\put(851,542){\makebox(0,0){$\diamondsuit$}}
\put(870,534){\makebox(0,0){$\diamondsuit$}}
\put(380,265){\makebox(0,0)[l]{$\sigma=0.25$}}
\put(384,599){\usebox{\plotpoint}}
\put(384,599){\usebox{\plotpoint}}
\put(404,596){\usebox{\plotpoint}}
\put(425,594){\usebox{\plotpoint}}
\put(445,592){\usebox{\plotpoint}}
\put(466,589){\usebox{\plotpoint}}
\put(487,587){\usebox{\plotpoint}}
\put(507,585){\usebox{\plotpoint}}
\put(528,582){\usebox{\plotpoint}}
\put(548,580){\usebox{\plotpoint}}
\put(569,578){\usebox{\plotpoint}}
\put(590,575){\usebox{\plotpoint}}
\put(610,573){\usebox{\plotpoint}}
\put(631,571){\usebox{\plotpoint}}
\put(652,568){\usebox{\plotpoint}}
\put(672,566){\usebox{\plotpoint}}
\put(693,564){\usebox{\plotpoint}}
\put(713,561){\usebox{\plotpoint}}
\put(734,559){\usebox{\plotpoint}}
\put(755,557){\usebox{\plotpoint}}
\put(775,554){\usebox{\plotpoint}}
\put(796,552){\usebox{\plotpoint}}
\put(816,546){\usebox{\plotpoint}}
\put(836,539){\usebox{\plotpoint}}
\put(855,533){\usebox{\plotpoint}}
\put(875,526){\usebox{\plotpoint}}
\put(890,522){\usebox{\plotpoint}}
\put(328,265){\raisebox{-1.2pt}{\makebox(0,0){$\heartsuit$}}}
\put(384,599){\raisebox{-1.2pt}{\makebox(0,0){$\heartsuit$}}}
\put(800,552){\raisebox{-1.2pt}{\makebox(0,0){$\heartsuit$}}}
\put(890,522){\raisebox{-1.2pt}{\makebox(0,0){$\heartsuit$}}}
\put(380,220){\makebox(0,0)[l]{$\sigma=0.0$}}
\put(384,584){\usebox{\plotpoint}}
\put(384,584){\usebox{\plotpoint}}
\put(404,583){\usebox{\plotpoint}}
\put(425,582){\usebox{\plotpoint}}
\put(446,581){\usebox{\plotpoint}}
\put(466,580){\usebox{\plotpoint}}
\put(487,579){\usebox{\plotpoint}}
\put(508,579){\usebox{\plotpoint}}
\put(529,578){\usebox{\plotpoint}}
\put(549,577){\usebox{\plotpoint}}
\put(570,575){\usebox{\plotpoint}}
\put(591,572){\usebox{\plotpoint}}
\put(611,569){\usebox{\plotpoint}}
\put(632,566){\usebox{\plotpoint}}
\put(652,564){\usebox{\plotpoint}}
\put(673,562){\usebox{\plotpoint}}
\put(694,560){\usebox{\plotpoint}}
\put(714,556){\usebox{\plotpoint}}
\put(734,551){\usebox{\plotpoint}}
\put(754,544){\usebox{\plotpoint}}
\put(773,537){\usebox{\plotpoint}}
\put(793,532){\usebox{\plotpoint}}
\put(814,528){\usebox{\plotpoint}}
\put(833,522){\usebox{\plotpoint}}
\put(852,514){\usebox{\plotpoint}}
\put(872,507){\usebox{\plotpoint}}
\put(891,499){\usebox{\plotpoint}}
\put(909,488){\usebox{\plotpoint}}
\put(919,483){\usebox{\plotpoint}}
\put(328,220){\makebox(0,0){$\clubsuit$}}
\put(384,584){\makebox(0,0){$\clubsuit$}}
\put(561,577){\makebox(0,0){$\clubsuit$}}
\put(634,566){\makebox(0,0){$\clubsuit$}}
\put(691,561){\makebox(0,0){$\clubsuit$}}
\put(738,551){\makebox(0,0){$\clubsuit$}}
\put(780,535){\makebox(0,0){$\clubsuit$}}
\put(819,528){\makebox(0,0){$\clubsuit$}}
\put(854,514){\makebox(0,0){$\clubsuit$}}
\put(887,502){\makebox(0,0){$\clubsuit$}}
\put(919,483){\makebox(0,0){$\clubsuit$}}
\end{picture}
{ \small
  \hspace{0.2cm}
  Figure~4: Average deficit angle $\langle \delta_t \rangle$ as a
  function of
  $\ell$ for different measure parameters $\sigma$. Remarkably,
  $\langle \delta_t \rangle$ stays always negative even across
  the transition to positive curvature.  The curves lie close together
  for $0 \leq \sigma \leq 1$ and differ significantly for $\sigma = 1.5$.
}
\end{figure}

The dependence of the simplicial path integral on the measure
within the framework of dynamical
triangulation has been studied by Br\"ugmann \cite{bru}.
By means of an additional term $n\sum_v\mbox{ln}[o(v)]$
in the action the type of the measure is
controlled via the parameter $n$.
Although the correspondence with the continuum is not entirely clear,
it is plausible that the cases with $n=-5$ and $n=0$
reproduce the scale-invariant and the uniform measure, respectively.
Besides a shift in $\ell$ a comparison\footnote{We use
$\ell=\mbox{sgn}(\beta)\sqrt{\frac{1}{2}|\beta|}$ as a
definition of the lattice spacing whenever $\ell<0$.}
of Figures~5 and 1 gives a remarkable
qualitative coincidence for both methods.

All these results show that the considered family of measures falls
into two qualitatively different classes \mbox{$0\le\sigma<1$} and
$\sigma>1$. The behavior of the considered expectation values suggests
that uniform and scale-invariant measure seem to belong to the same
class.

\begin{figure}[t]
\setlength{\unitlength}{0.240900pt}
\ifx\plotpoint\undefined\newsavebox{\plotpoint}\fi
\sbox{\plotpoint}{\rule[-0.175pt]{0.350pt}{0.350pt}}%
\begin{picture}(1050,850)(60,0)
\tenrm
\sbox{\plotpoint}{\rule[-0.175pt]{0.350pt}{0.350pt}}%
\put(264,263){\rule[-0.175pt]{173.689pt}{0.350pt}}
\put(264,158){\rule[-0.175pt]{4.818pt}{0.350pt}}
\put(242,158){\makebox(0,0)[r]{-10}}
\put(965,158){\rule[-0.175pt]{4.818pt}{0.350pt}}
\put(264,263){\rule[-0.175pt]{4.818pt}{0.350pt}}
\put(242,263){\makebox(0,0)[r]{0}}
\put(965,263){\rule[-0.175pt]{4.818pt}{0.350pt}}
\put(264,368){\rule[-0.175pt]{4.818pt}{0.350pt}}
\put(242,368){\makebox(0,0)[r]{10}}
\put(965,368){\rule[-0.175pt]{4.818pt}{0.350pt}}
\put(264,473){\rule[-0.175pt]{4.818pt}{0.350pt}}
\put(242,473){\makebox(0,0)[r]{20}}
\put(965,473){\rule[-0.175pt]{4.818pt}{0.350pt}}
\put(264,577){\rule[-0.175pt]{4.818pt}{0.350pt}}
\put(242,577){\makebox(0,0)[r]{30}}
\put(965,577){\rule[-0.175pt]{4.818pt}{0.350pt}}
\put(264,682){\rule[-0.175pt]{4.818pt}{0.350pt}}
\put(242,682){\makebox(0,0)[r]{40}}
\put(965,682){\rule[-0.175pt]{4.818pt}{0.350pt}}
\put(264,787){\rule[-0.175pt]{4.818pt}{0.350pt}}
\put(242,787){\makebox(0,0)[r]{50}}
\put(965,787){\rule[-0.175pt]{4.818pt}{0.350pt}}
\put(336,158){\rule[-0.175pt]{0.350pt}{4.818pt}}
\put(336,113){\makebox(0,0){-1}}
\put(336,767){\rule[-0.175pt]{0.350pt}{4.818pt}}
\put(480,158){\rule[-0.175pt]{0.350pt}{4.818pt}}
\put(480,113){\makebox(0,0){-0.5}}
\put(480,767){\rule[-0.175pt]{0.350pt}{4.818pt}}
\put(625,158){\rule[-0.175pt]{0.350pt}{4.818pt}}
\put(625,113){\makebox(0,0){0}}
\put(625,767){\rule[-0.175pt]{0.350pt}{4.818pt}}
\put(769,158){\rule[-0.175pt]{0.350pt}{4.818pt}}
\put(769,113){\makebox(0,0){0.5}}
\put(769,767){\rule[-0.175pt]{0.350pt}{4.818pt}}
\put(913,158){\rule[-0.175pt]{0.350pt}{4.818pt}}
\put(913,113){\makebox(0,0){1}}
\put(913,767){\rule[-0.175pt]{0.350pt}{4.818pt}}
\put(264,158){\rule[-0.175pt]{173.689pt}{0.350pt}}
\put(985,158){\rule[-0.175pt]{0.350pt}{151.526pt}}
\put(264,787){\rule[-0.175pt]{173.689pt}{0.350pt}}
\put(90,472){\makebox(0,0)[l]{\shortstack{$\langle\tilde R\rangle$}}}
\put(624,40){\makebox(0,0){$\ell$}}
\put(264,158){\rule[-0.175pt]{0.350pt}{151.526pt}}
\sbox{\plotpoint}{\rule[-0.250pt]{0.500pt}{0.500pt}}%
\put(370,737){\makebox(0,0)[l]{$n=5$}}
\put(927,754){\usebox{\plotpoint}}
\put(927,754){\usebox{\plotpoint}}
\put(906,751){\usebox{\plotpoint}}
\put(887,743){\usebox{\plotpoint}}
\put(870,733){\usebox{\plotpoint}}
\put(859,715){\usebox{\plotpoint}}
\put(848,697){\usebox{\plotpoint}}
\put(838,679){\usebox{\plotpoint}}
\put(830,660){\usebox{\plotpoint}}
\put(822,641){\usebox{\plotpoint}}
\put(815,621){\usebox{\plotpoint}}
\put(807,602){\usebox{\plotpoint}}
\put(799,583){\usebox{\plotpoint}}
\put(793,563){\usebox{\plotpoint}}
\put(786,543){\usebox{\plotpoint}}
\put(780,524){\usebox{\plotpoint}}
\put(773,504){\usebox{\plotpoint}}
\put(767,484){\usebox{\plotpoint}}
\put(760,465){\usebox{\plotpoint}}
\put(754,445){\usebox{\plotpoint}}
\put(747,425){\usebox{\plotpoint}}
\put(734,409){\usebox{\plotpoint}}
\put(721,393){\usebox{\plotpoint}}
\put(708,376){\usebox{\plotpoint}}
\put(696,360){\usebox{\plotpoint}}
\put(683,344){\usebox{\plotpoint}}
\put(670,327){\usebox{\plotpoint}}
\put(657,311){\usebox{\plotpoint}}
\put(645,294){\usebox{\plotpoint}}
\put(632,278){\usebox{\plotpoint}}
\put(617,265){\usebox{\plotpoint}}
\put(598,255){\usebox{\plotpoint}}
\put(580,246){\usebox{\plotpoint}}
\put(561,236){\usebox{\plotpoint}}
\put(543,227){\usebox{\plotpoint}}
\put(524,218){\usebox{\plotpoint}}
\put(506,208){\usebox{\plotpoint}}
\put(486,202){\usebox{\plotpoint}}
\put(466,197){\usebox{\plotpoint}}
\put(445,193){\usebox{\plotpoint}}
\put(425,192){\usebox{\plotpoint}}
\put(404,191){\usebox{\plotpoint}}
\put(383,191){\usebox{\plotpoint}}
\put(363,191){\usebox{\plotpoint}}
\put(342,191){\usebox{\plotpoint}}
\put(322,191){\usebox{\plotpoint}}
\put(320,737){\makebox(0,0){$\bigtriangledown$}}
\put(927,754){\makebox(0,0){$\triangle$}}
\put(901,751){\makebox(0,0){$\triangle$}}
\put(872,736){\makebox(0,0){$\triangle$}}
\put(839,682){\makebox(0,0){$\triangle$}}
\put(799,581){\makebox(0,0){$\triangle$}}
\put(748,427){\makebox(0,0){$\triangle$}}
\put(625,269){\makebox(0,0){$\triangle$}}
\put(501,206){\makebox(0,0){$\triangle$}}
\put(450,194){\makebox(0,0){$\triangle$}}
\put(410,192){\makebox(0,0){$\triangle$}}
\put(377,191){\makebox(0,0){$\triangle$}}
\put(348,192){\makebox(0,0){$\triangle$}}
\put(322,191){\makebox(0,0){$\triangle$}}
\put(370,692){\makebox(0,0)[l]{$n=1$}}
\put(927,756){\usebox{\plotpoint}}
\put(927,756){\usebox{\plotpoint}}
\put(906,754){\usebox{\plotpoint}}
\put(885,752){\usebox{\plotpoint}}
\put(865,748){\usebox{\plotpoint}}
\put(845,742){\usebox{\plotpoint}}
\put(829,730){\usebox{\plotpoint}}
\put(814,715){\usebox{\plotpoint}}
\put(799,700){\usebox{\plotpoint}}
\put(794,681){\usebox{\plotpoint}}
\put(789,660){\usebox{\plotpoint}}
\put(784,640){\usebox{\plotpoint}}
\put(780,620){\usebox{\plotpoint}}
\put(775,600){\usebox{\plotpoint}}
\put(770,580){\usebox{\plotpoint}}
\put(765,559){\usebox{\plotpoint}}
\put(760,539){\usebox{\plotpoint}}
\put(755,519){\usebox{\plotpoint}}
\put(751,499){\usebox{\plotpoint}}
\put(744,479){\usebox{\plotpoint}}
\put(733,462){\usebox{\plotpoint}}
\put(723,444){\usebox{\plotpoint}}
\put(712,426){\usebox{\plotpoint}}
\put(701,408){\usebox{\plotpoint}}
\put(690,390){\usebox{\plotpoint}}
\put(680,373){\usebox{\plotpoint}}
\put(669,355){\usebox{\plotpoint}}
\put(658,337){\usebox{\plotpoint}}
\put(648,319){\usebox{\plotpoint}}
\put(637,302){\usebox{\plotpoint}}
\put(626,284){\usebox{\plotpoint}}
\put(609,273){\usebox{\plotpoint}}
\put(591,263){\usebox{\plotpoint}}
\put(572,254){\usebox{\plotpoint}}
\put(554,244){\usebox{\plotpoint}}
\put(536,235){\usebox{\plotpoint}}
\put(517,225){\usebox{\plotpoint}}
\put(499,216){\usebox{\plotpoint}}
\put(479,210){\usebox{\plotpoint}}
\put(459,204){\usebox{\plotpoint}}
\put(438,201){\usebox{\plotpoint}}
\put(418,199){\usebox{\plotpoint}}
\put(397,198){\usebox{\plotpoint}}
\put(376,197){\usebox{\plotpoint}}
\put(356,197){\usebox{\plotpoint}}
\put(335,197){\usebox{\plotpoint}}
\put(322,197){\usebox{\plotpoint}}
\put(320,692){\raisebox{-1.2pt}{\makebox(0,0){$\Box$}}}
\put(927,756){\raisebox{-1.2pt}{\makebox(0,0){$\Diamond$}}}
\put(901,754){\raisebox{-1.2pt}{\makebox(0,0){$\Diamond$}}}
\put(872,751){\raisebox{-1.2pt}{\makebox(0,0){$\Diamond$}}}
\put(839,740){\raisebox{-1.2pt}{\makebox(0,0){$\Diamond$}}}
\put(799,700){\raisebox{-1.2pt}{\makebox(0,0){$\Diamond$}}}
\put(748,486){\raisebox{-1.2pt}{\makebox(0,0){$\Diamond$}}}
\put(625,281){\raisebox{-1.2pt}{\makebox(0,0){$\Diamond$}}}
\put(501,217){\raisebox{-1.2pt}{\makebox(0,0){$\Diamond$}}}
\put(450,202){\raisebox{-1.2pt}{\makebox(0,0){$\Diamond$}}}
\put(410,199){\raisebox{-1.2pt}{\makebox(0,0){$\Diamond$}}}
\put(377,197){\raisebox{-1.2pt}{\makebox(0,0){$\Diamond$}}}
\put(348,197){\raisebox{-1.2pt}{\makebox(0,0){$\Diamond$}}}
\put(322,197){\raisebox{-1.2pt}{\makebox(0,0){$\Diamond$}}}
\put(370,647){\makebox(0,0)[l]{$n=0$}}
\put(927,756){\usebox{\plotpoint}}
\put(927,756){\usebox{\plotpoint}}
\put(906,755){\usebox{\plotpoint}}
\put(885,752){\usebox{\plotpoint}}
\put(865,749){\usebox{\plotpoint}}
\put(845,743){\usebox{\plotpoint}}
\put(827,734){\usebox{\plotpoint}}
\put(809,722){\usebox{\plotpoint}}
\put(795,708){\usebox{\plotpoint}}
\put(787,689){\usebox{\plotpoint}}
\put(779,670){\usebox{\plotpoint}}
\put(771,651){\usebox{\plotpoint}}
\put(762,631){\usebox{\plotpoint}}
\put(754,612){\usebox{\plotpoint}}
\put(746,593){\usebox{\plotpoint}}
\put(738,574){\usebox{\plotpoint}}
\put(730,555){\usebox{\plotpoint}}
\put(722,536){\usebox{\plotpoint}}
\put(714,517){\usebox{\plotpoint}}
\put(706,497){\usebox{\plotpoint}}
\put(699,478){\usebox{\plotpoint}}
\put(691,459){\usebox{\plotpoint}}
\put(683,440){\usebox{\plotpoint}}
\put(675,421){\usebox{\plotpoint}}
\put(667,401){\usebox{\plotpoint}}
\put(659,382){\usebox{\plotpoint}}
\put(651,363){\usebox{\plotpoint}}
\put(643,344){\usebox{\plotpoint}}
\put(636,325){\usebox{\plotpoint}}
\put(628,305){\usebox{\plotpoint}}
\put(614,291){\usebox{\plotpoint}}
\put(596,281){\usebox{\plotpoint}}
\put(578,270){\usebox{\plotpoint}}
\put(560,260){\usebox{\plotpoint}}
\put(542,249){\usebox{\plotpoint}}
\put(524,239){\usebox{\plotpoint}}
\put(507,228){\usebox{\plotpoint}}
\put(488,220){\usebox{\plotpoint}}
\put(468,213){\usebox{\plotpoint}}
\put(448,206){\usebox{\plotpoint}}
\put(428,203){\usebox{\plotpoint}}
\put(407,200){\usebox{\plotpoint}}
\put(387,200){\usebox{\plotpoint}}
\put(366,199){\usebox{\plotpoint}}
\put(345,199){\usebox{\plotpoint}}
\put(324,199){\usebox{\plotpoint}}
\put(322,199){\usebox{\plotpoint}}
\put(320,647){\circle{24}}
\put(927,756){\circle{24}}
\put(901,755){\circle{24}}
\put(872,751){\circle{24}}
\put(839,742){\circle{24}}
\put(799,715){\circle{24}}
\put(748,598){\circle{24}}
\put(625,298){\circle{24}}
\put(501,225){\circle{24}}
\put(450,207){\circle{24}}
\put(410,201){\circle{24}}
\put(377,200){\circle{24}}
\put(348,199){\circle{24}}
\put(322,199){\circle{24}}
\put(370,602){\makebox(0,0)[l]{$n=-1$}}
\put(927,756){\usebox{\plotpoint}}
\put(927,756){\usebox{\plotpoint}}
\put(906,755){\usebox{\plotpoint}}
\put(885,752){\usebox{\plotpoint}}
\put(865,749){\usebox{\plotpoint}}
\put(845,743){\usebox{\plotpoint}}
\put(826,734){\usebox{\plotpoint}}
\put(808,723){\usebox{\plotpoint}}
\put(793,710){\usebox{\plotpoint}}
\put(780,694){\usebox{\plotpoint}}
\put(768,677){\usebox{\plotpoint}}
\put(755,661){\usebox{\plotpoint}}
\put(744,643){\usebox{\plotpoint}}
\put(737,624){\usebox{\plotpoint}}
\put(729,605){\usebox{\plotpoint}}
\put(721,585){\usebox{\plotpoint}}
\put(713,566){\usebox{\plotpoint}}
\put(705,547){\usebox{\plotpoint}}
\put(697,528){\usebox{\plotpoint}}
\put(690,509){\usebox{\plotpoint}}
\put(682,489){\usebox{\plotpoint}}
\put(674,470){\usebox{\plotpoint}}
\put(666,451){\usebox{\plotpoint}}
\put(658,432){\usebox{\plotpoint}}
\put(650,413){\usebox{\plotpoint}}
\put(642,393){\usebox{\plotpoint}}
\put(635,374){\usebox{\plotpoint}}
\put(627,355){\usebox{\plotpoint}}
\put(614,339){\usebox{\plotpoint}}
\put(599,325){\usebox{\plotpoint}}
\put(584,311){\usebox{\plotpoint}}
\put(569,296){\usebox{\plotpoint}}
\put(554,282){\usebox{\plotpoint}}
\put(539,268){\usebox{\plotpoint}}
\put(524,253){\usebox{\plotpoint}}
\put(508,239){\usebox{\plotpoint}}
\put(491,228){\usebox{\plotpoint}}
\put(472,220){\usebox{\plotpoint}}
\put(453,213){\usebox{\plotpoint}}
\put(433,209){\usebox{\plotpoint}}
\put(412,205){\usebox{\plotpoint}}
\put(391,203){\usebox{\plotpoint}}
\put(371,201){\usebox{\plotpoint}}
\put(350,201){\usebox{\plotpoint}}
\put(329,201){\usebox{\plotpoint}}
\put(322,201){\usebox{\plotpoint}}
\put(320,602){\raisebox{-1.2pt}{\makebox(0,0){$\Diamond$}}}
\put(927,756){\raisebox{-1.2pt}{\makebox(0,0){$\Box$}}}
\put(901,755){\raisebox{-1.2pt}{\makebox(0,0){$\Box$}}}
\put(872,751){\raisebox{-1.2pt}{\makebox(0,0){$\Box$}}}
\put(839,742){\raisebox{-1.2pt}{\makebox(0,0){$\Box$}}}
\put(799,718){\raisebox{-1.2pt}{\makebox(0,0){$\Box$}}}
\put(748,651){\raisebox{-1.2pt}{\makebox(0,0){$\Box$}}}
\put(625,350){\raisebox{-1.2pt}{\makebox(0,0){$\Box$}}}
\put(501,232){\raisebox{-1.2pt}{\makebox(0,0){$\Box$}}}
\put(450,212){\raisebox{-1.2pt}{\makebox(0,0){$\Box$}}}
\put(410,205){\raisebox{-1.2pt}{\makebox(0,0){$\Box$}}}
\put(377,202){\raisebox{-1.2pt}{\makebox(0,0){$\Box$}}}
\put(348,201){\raisebox{-1.2pt}{\makebox(0,0){$\Box$}}}
\put(322,201){\raisebox{-1.2pt}{\makebox(0,0){$\Box$}}}
\put(370,557){\makebox(0,0)[l]{$n=-5$}}
\put(927,756){\usebox{\plotpoint}}
\put(927,756){\usebox{\plotpoint}}
\put(906,755){\usebox{\plotpoint}}
\put(885,753){\usebox{\plotpoint}}
\put(865,750){\usebox{\plotpoint}}
\put(844,746){\usebox{\plotpoint}}
\put(825,739){\usebox{\plotpoint}}
\put(806,731){\usebox{\plotpoint}}
\put(788,720){\usebox{\plotpoint}}
\put(771,708){\usebox{\plotpoint}}
\put(754,696){\usebox{\plotpoint}}
\put(736,685){\usebox{\plotpoint}}
\put(718,676){\usebox{\plotpoint}}
\put(700,666){\usebox{\plotpoint}}
\put(681,656){\usebox{\plotpoint}}
\put(663,646){\usebox{\plotpoint}}
\put(645,636){\usebox{\plotpoint}}
\put(627,627){\usebox{\plotpoint}}
\put(610,614){\usebox{\plotpoint}}
\put(594,600){\usebox{\plotpoint}}
\put(579,587){\usebox{\plotpoint}}
\put(563,574){\usebox{\plotpoint}}
\put(547,560){\usebox{\plotpoint}}
\put(531,547){\usebox{\plotpoint}}
\put(515,534){\usebox{\plotpoint}}
\put(500,520){\usebox{\plotpoint}}
\put(491,501){\usebox{\plotpoint}}
\put(483,482){\usebox{\plotpoint}}
\put(475,463){\usebox{\plotpoint}}
\put(467,444){\usebox{\plotpoint}}
\put(458,425){\usebox{\plotpoint}}
\put(450,406){\usebox{\plotpoint}}
\put(441,387){\usebox{\plotpoint}}
\put(433,368){\usebox{\plotpoint}}
\put(424,349){\usebox{\plotpoint}}
\put(416,330){\usebox{\plotpoint}}
\put(406,312){\usebox{\plotpoint}}
\put(394,295){\usebox{\plotpoint}}
\put(382,278){\usebox{\plotpoint}}
\put(367,264){\usebox{\plotpoint}}
\put(350,253){\usebox{\plotpoint}}
\put(331,245){\usebox{\plotpoint}}
\put(322,242){\usebox{\plotpoint}}
\put(320,555){\makebox(0,0){$\triangle$}}
\put(927,756){\makebox(0,0){$\bigtriangledown$}}
\put(901,755){\makebox(0,0){$\bigtriangledown$}}
\put(872,752){\makebox(0,0){$\bigtriangledown$}}
\put(839,745){\makebox(0,0){$\bigtriangledown$}}
\put(799,728){\makebox(0,0){$\bigtriangledown$}}
\put(748,692){\makebox(0,0){$\bigtriangledown$}}
\put(625,626){\makebox(0,0){$\bigtriangledown$}}
\put(501,522){\makebox(0,0){$\bigtriangledown$}}
\put(450,405){\makebox(0,0){$\bigtriangledown$}}
\put(410,317){\makebox(0,0){$\bigtriangledown$}}
\put(377,270){\makebox(0,0){$\bigtriangledown$}}
\put(348,252){\makebox(0,0){$\bigtriangledown$}}
\put(322,242){\makebox(0,0){$\bigtriangledown$}}
\end{picture}
{ \small
  \hspace{0.2cm}
  Figure~5: Investigations of different measures with dynamical triangulation
  of the 4-sphere by Br\"ugmann \cite{bru}. An additional term
  in the action mimics different
  measures. Besides a shift in $\ell$ the picture
  has a striking similarity to Figure~1 if one identifies $\sigma = 0$
  with $n = -5$ (scale-invariant measure) and
  $\sigma = 1$ with $n = 0$ (uniform measure).
  Notice the small influence of $n$ in the range $-5 \leq n \leq +1$ and
  the exceptional behavior of $n = +5$.
  }
\end{figure}

\end{document}